\documentclass[twocolumn,amsmath,amssymb,prl,aps,superscriptaddress]{revtex4-2}
\usepackage[version=3]{mhchem} % Formula subscripts using \ce{}
\usepackage{amsmath}
\usepackage{upgreek}
\usepackage{color}
\usepackage{textgreek}
\usepackage{graphicx,subfigure}% Include figure files
\usepackage{hyperref}
\usepackage{nicefrac}
\usepackage{dcolumn}% Align table columns on decimal point
\usepackage{bm}% bold math
\usepackage[normalem]{ulem}
\usepackage{xcolor}

\usepackage{xspace}

\begin{document}

\title{Motility and self-organization of gliding \textit{Chlamydomonas} populations}

\author{Sebastian Till}
\affiliation{Max Planck Institute for Dynamics and Self-Organization (MPIDS), Am Fa{\ss}berg 17, D-37077 G\"{o}ttingen, Germany}

\author{Florian Ebmeier}
\affiliation{Max Planck Institute for Dynamics and Self-Organization (MPIDS), Am Fa{\ss}berg 17, D-37077 G\"{o}ttingen, Germany}
\affiliation{Department of Computer Science, Experimental Cognitive Science, University of T\"{u}bingen, Sand 13, D-72076 T\"{u}bingen, Germany}

\author{Alexandros A.\ Fragkopoulos}
\affiliation{Max Planck Institute for Dynamics and Self-Organization (MPIDS), Am Fa{\ss}berg 17, D-37077 G\"{o}ttingen, Germany}
\affiliation{Experimental Physics V, University of Bayreuth, Universit\"atsstr.\ 30, D-95447 Bayreuth, Germany}

\author{Marco G.\ Mazza}
\thanks{E-mail: m.g.mazza@lboro.ac.uk}
\affiliation{Max Planck Institute for Dynamics and Self-Organization (MPIDS), Am Fa{\ss}berg 17, D-37077 G\"{o}ttingen, Germany}
\affiliation{Interdisciplinary Centre for Mathematical Modelling and Department of Mathematical Sciences, Loughborough University, Loughborough, Leicestershire LE11 3TU, United Kingdom}

\author{Oliver B\"{a}umchen}
\thanks{E-mail: oliver.baeumchen@uni-bayreuth.de}
\affiliation{Max Planck Institute for Dynamics and Self-Organization (MPIDS), Am Fa{\ss}berg 17, D-37077 G\"{o}ttingen, Germany}
\affiliation{Experimental Physics V, University of Bayreuth, Universit\"atsstr.\ 30, D-95447 Bayreuth, Germany}

\begin{abstract} 
\noindent Cellular appendages such as cilia and flagella represent universal tools enabling cells and microbes, among other essential functionalities, to propel themselves in diverse environments. 
In its planktonic, i.e.\ freely swimming, state the unicellular bi-flagellated microbe \textit{Chlamydomonas reinhardtii} employs a periodic breaststroke-like flagellar beating to displace the surrounding fluid.
Another flagella-mediated motility mode is observed for surface-associated \textit{Chlamydomonas} cells, which glide along the surface by means of force transduction through an intraflagellar transport machinery.
Experiments and statistical motility analysis demonstrate that this gliding motility enhances clustering and supports self-organization of \textit{Chlamydomonas} populations.
We employ Minkowski functionals to characterize the spatiotemporal organization of the surface-associated cell monolayer.
We find that simulations based on a purely mechanistic approach cannot capture the observed non-random cell configurations. 
Quantitative agreement with experimental data however is achieved when considering a minimal cognitive model of the flagellar mechanosensing. 

\end{abstract}

\maketitle

Motility is a key feature of microorganisms to respond to environmental cues and to actively search for favourable living conditions, nutrient sources and mating partners \cite{Berg1973,Miller2001,Lauga_2009}.
Microbial self-propulsion can be realized by means of shape deformations \cite{Schulman2014,Noselli2019}, the formation of lamellipodia \cite{Mitchinson1996,Batchelder2011} and the periodic actuation of single or multiple cilia or flagella \cite{Rossi13085,Wan2018,Tsang2018}.
Such cellular appendages represent universal building blocks of life that enable cells and microbes to sense and interact with their environment.
The fast and coordinated actuation of the two flagella of \textit{Chlamydomonas reinhardtii} \cite{Ruffer1985,drescher2010,friedrich2012,geyer2013,wan2016coordinated,Boeddeker2020}, which is capable of propelling the cell body in a liquid medium \cite{polin2009,ostapenko2018curvature}, has received ample attention recently as a prime model system.
\textit{Chlamydomonas} dwells in complex geometric confinement \cite{harris2009chlamydomonas} and exhibits light-regulated and flagella-mediated adhesion to surfaces, i.e.\ the cells may transition from the free-swimming state to a surface-associated state \cite{kreis2018adhesion,Kreis2019}.
Even in the latter state, the cells are not static: an intraflagellar transport (IFT) machinery \cite{Stepanek2016} translocates the cell body along the flagella \cite{Shih2013}, a process termed gliding motility \cite{Bloodgood1981,Xu2020}.
However, the purpose of this gliding motility to date \textit{`still remains a mystery'} \cite{harris2009chlamydomonas}.

\begin{figure}[b]
    \centering
    \includegraphics{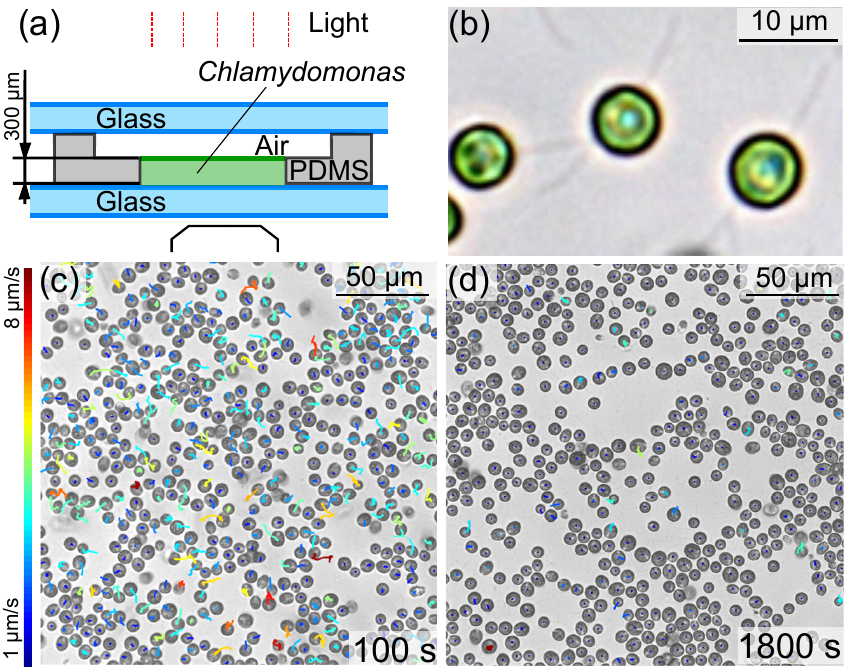}
    \caption{(a) Experimental setup: A \textit{Chlamydomonas} suspension is confined between glass and air. A second glass slide seals the suspension to prevent evaporation. The cells adhere to the bottom glass slide when illuminated with blue light (top illumination). (b) Upon adhesion \textit{Chlamydomonas} flagella orient in the 180$^\circ$ gliding configuration. (c),(d) Snapshots of the cell population at early ($t$=100\,s) and late times ($t$=1800\,s). Colored tracks represent cell displacements over 5\,s. Over time the cells form large-scale structures.}
    \label{fig:FIG1}
\end{figure}

%%%%%%%%%%%%%%%
\begin{figure}[b]
    \centering
    \includegraphics{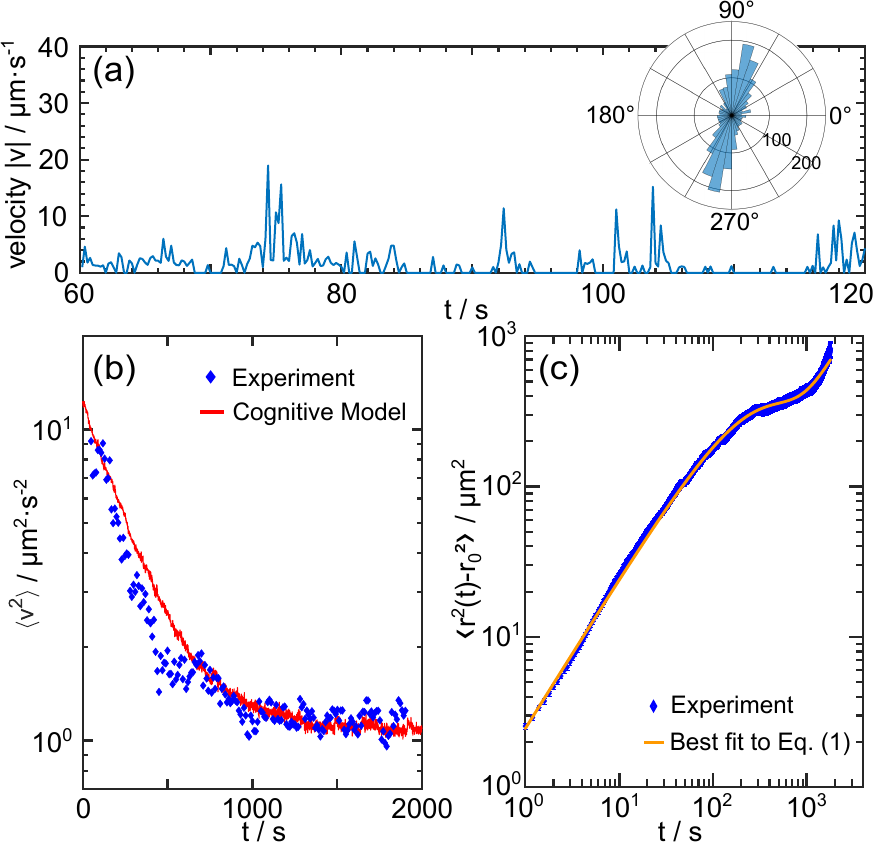}
    \caption{(a) The motion of a single cell is characterised by abrupt displacements followed by periods of diminished activity. 
    The gliding configuration of the flagella induce a preferred direction of motion, as shown for the measured orientation of one cell over 2663 frames. (see inset). 
    (b) Experimental (blue diamonds) and cognitive-model simulated (red line) mean-squared velocity at a filling fraction of $\Phi=22\%$.
    (c) Population-averaged mean squared displacement of the cells (blue diamonds) and best fit (orange line) to Eq.~(\ref{eq:eq1}) at a filling fraction of $\Phi=32\%$.
    }
    \label{fig:FIG2}
\end{figure}
%%%%%%%%%%%%%%%

In this \textit{Letter}, we demonstrate that gliding motility enables surface-associated \textit{Chlamydomonas} cells to cluster and form compact, interconnected microbial communities.
We analyze the statistics of cell trajectories and characterize the spatiotemporal evolution of the cell positions within the population using two-dimensional Minkowski functionals.
Simulations successfully capture the non-random cell positions for different cell densities from very dilute systems of merely isolated cells to densely-packed monolayers.
These simulations go well beyond a purely mechanistic approach, which we show fails at capturing the experimental data, and include cognitive forces to recover the spatiotemporal dynamics of the cell population.

\textit{-- Experiments} Experiments are performed using suspensions of motile cells confined in a liquid film that is supported by a glass surface, on which the cells may adhere upon a switch from red to blue light \cite{kreis2018adhesion}, see Fig.~\ref{fig:FIG1}(a).
We use the wild-type \textit{Chlamydomonas reinhardtii} strain SAG11-32b, which we cultivate axenically in Tris-Acetate-Phosphate (TAP) medium on a 12h/12h day/night cycle following established recipes \cite{kreis2018adhesion}.
The surface-adhered cells are observed using bright-field video microscopy in inverted configuration (Olympus IX-83) while being illuminated from the top using narrow bandpass filters for red ($\lambda=671\pm 6\,$nm) light before adsorption and blue light ($\lambda=470\pm 6\,$nm) to induce surface adhesion.
After the surface association process, most cells have achieved the gliding configuration with both flagella in a widespread 180$^\circ$ configuration, see Fig.~\ref{fig:FIG1}(b).
Time-resolved cell positions are recorded using a monochromatic camera (FLIR Systems, GS3-U3-41C6M-C, 2048x2048 pixels) at 5 frames per second and analyzed using digital image processing following established cell tracking protocols \footnote{Cell tracking code is available at \url{https://doi.org/10.5281/zenodo.4449791}.}.
We find that the population exhibits a relatively high activity during which cells adsorb to the glass substrate and eventually a plateau cell density is achieved, see Fig.~\ref{fig:FIG1}(c). 
Using the displacement of the cells, we observe that this overall activity decays over time and the cells reach quasi-static positions, which differ from their initial adsorption sites and show distinct signatures of clustering and compactification, see Fig.~\ref{fig:FIG1}(d). 
We now dissect the link between cell motility and clustering by a) characterizing the motility signatures on the single-cell as well as on the population level and b) analyzing the cell positions from a morphological perspective.

%%%%%%%%%%%%%%%
\begin{figure*}[t]
    \centering
    \includegraphics{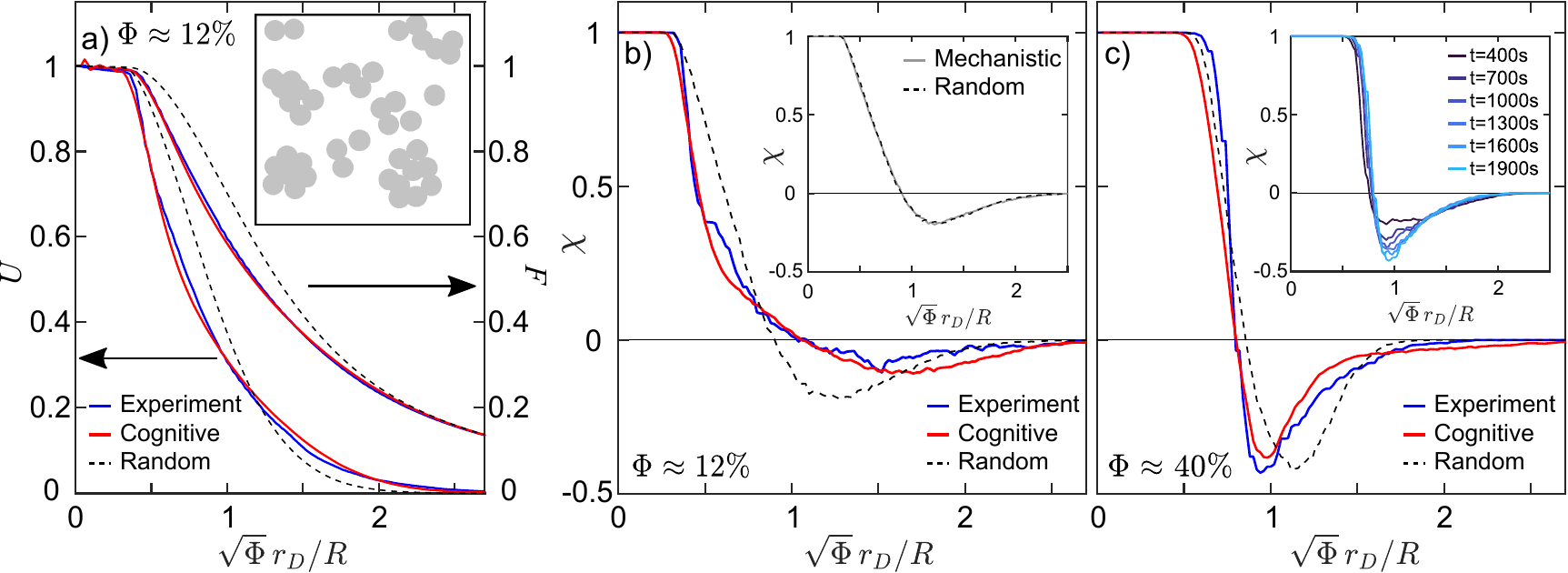}
    \caption{(a) Normalized area $F$ and boundary length $U$ at $\Phi=12\%$ and (b,c) normalized Euler number $\chi$ at (b) $\Phi=12\%$ and (c) $\Phi=40\%$ for experimental data (blue line), cognitive-model simulated data (red line), and randomly distributed objects (black line).
    The position of cells (cell radius $R$) are dilated to discs with radius of dilation $r_\mathrm{D}$ (inset in (a)).
    The mechanistic model reproduces a random distribution and it is insufficient to capture the experimental data (inset in (b)).
    Experimental data at $\Phi=40\%$ are displayed at different times during the experiment (inset in (c)). 
    (b) For low filling fractions, $\chi$ exhibits a distinct slope indicating local clustering. 
    (c) For high filling fractions, it exhibits a sudden decrease at low $r_D$ indicating that above the percolation threshold ($\Phi\geq 38\%$) there is a single global cluster.
    }
    \label{fig:FIG3}
\end{figure*}
%%%%%%%%%%%%%%%

\textit{-- Motility Analysis} The biological process of gliding motility is based on a set of glycoproteins that mediate the adhesion between the surface and the flagella membrane \cite{Bloodgood1981,Shih2013}. However, these adhesion sites are attached to IFT trains and can be translocated through molecular motors towards the cell body, effectively pulling the cell body towards adhesion sites \cite{Shih2013}. Since these forces are applied on both flagella, the cell can experience a stochastic force in either direction. In addition, occasional elevation of Ca$^{+2}$ on one of the flagellum, transiently reverses the direction of the IFT trains causing the cell to move rapidly away from the activated flagellum \cite{collingridge2013compartmentalized,fort2021ca2+}.

We start with single-cell tracking, and extract position, velocity and directionality of the cell's motion.
We find that the cell velocity exhibits distinct bursts of activity followed by pause periods, see Fig.~\ref{fig:FIG2}(a), which is likely reminiscent to the tug-of-war of active IFT trains on both flagella \cite{Shih2013}; in the following we refer to these bursts of activity as `intermittency'.
As a result of the specific gliding configuration of the two flagella, the directionality of the motion is predominantly, but not exclusively, constrained to the initial angle of flagella orientation with respect to the lab frame, see inset of Fig.~\ref{fig:FIG2}(a).
On occasion, one flagellum might also transiently detach, which may result in reorientation events \cite{Shih2013}.
In order to characterize the spatiotemporal evolution and statistics of cell trajectories, we employ population-averaged quantitative measures. 
The mean-squared velocity (MSV) is found to decrease exponentially, exhibiting a cell density-independent decay time $\tau$ of about  390$\pm$50\,s with a maximum and minimum squared velocity of about $v^2_{max}=13\pm1$ \textmu {m}$^2$ {s}$^{-2}$ and $v^2_{min}=1.2\pm0.2$ \textmu {m}$^2$ {s}$^{-2}$, with the latter matching the speed of a single dynein motor \cite{kural2005kinesin}, see Fig.~\ref{fig:FIG2}(b).
In addition, we quantify the orientation of the cells using the orientation auto-correlation function (OACF), $\langle \mathbf{\hat e_\phi} (t_0)\cdot \mathbf{\hat e_\phi}(t_0+t)\rangle$, where $\mathbf{\hat e_\phi}$ is the orientation of the flagella \footnote{See Supplemental Material at [URL] for details of the theoretical models, simulation parameters and additional experimental data.}.
We find that the orientation decorrelates on characteristic time scales of tens of seconds, which is a result of transient flagella de- and reattachment events and reorientation during gliding \cite{Shih2013}.
Specifically, the OACF decays with a characteristic time $\tau_D$, which is measured as OACF$(\tau_D)=1/e$, and is related to the orientational diffusion constant $D_r=\tau_D^{-1}$ (see Fig.~S1 in \footnotemark[2]).
By calculating $\tau_D$ at different times $t_0$ of the experiment, we find that $\tau_D$ progressively increases, indicating that the cells rotate less over time.
By considering the loss of activity from the MSV and describing the single-cell motility as predominantly diffusive, a superposition of both mechanisms provides an analytical description of the time-dependent mean-squared displacement (MSD) as
\begin{equation}\label{eq:eq1}
    \langle r^2(t)\rangle=2 v^2\tau_\mathrm{T} t\,,\quad  v^2=A_\mathrm{1}\exp{(-t/\tau)}+A_\mathrm{0}\,,
\end{equation}
where $A_\mathrm{0}$ and $A_\mathrm{1}$ are constants describing the activity of the system, $\tau$  the decay time, and $\tau_\mathrm{T}$ the reversal time. 
We find that Eq.~\eqref{eq:eq1} %and (\ref{eq:eq1b}) 
describes the observed MSD, shown in Fig.~\ref{fig:FIG2}(c), with fitting parameters $A_{1}=4.2 \pm 0.6$ \textmu {m}$^2$ {s}$^{-2}$, $\tau=248\pm 26$ s, while $A_\mathrm{0}=v^2_{min}$ is set as a fixed parameter obtained from the MSV.

\textit{-- Morphological Analysis} We find that the two-point correlation function, which describe positional correlations between objects, is insensitive to our experimental observations (see Fig.~S2 in \footnotemark[2]). Hence, we turn towards an alternative morphological analysis of the spatiotemporal evolution of cell positions.
For different surface packing fractions, $\Phi$, we calculate the two-dimensional (2D) Minkowski functionals, i.e.\ area, boundary length and Euler number \cite{Mecke2000}.
Minkowski functionals are elegant tools from integral geometry (persistent homology) to dissect morphological information regarding the evolution of object positions in space and time and have been successfully applied to a plethora of systems, from nucleated holes in a liquid film \cite{Herminghaus1998}, colloidal suspensions \cite{Hoffmann2006}, and soil structure \cite{vogel2010} to galaxies in the universe \cite{Mecke1994}.
Detected cell centers are used as nuclei for inflating 2D discs with radius $r_\mathrm{D}$, see inset of Fig.~\ref{fig:FIG3}(a).
At different values of $r_\mathrm{D}$, the area, boundary length, and Euler number are calculated. The latter is defined as the difference between the number of connected regions and holes in a binary 2D image.
The normalized area $F$, boundary length $U$, and Euler number $\chi$ (normalized with $N(\pi r_D^2)$, $N(2\pi r_D)$, and $N$, respectively, where $N$ is the total number of cells) is shown in Fig.~\ref{fig:FIG3}(a,b) in case of an experimental cell packing fraction of about 12\%.
Upon inflating the discs the spatial connectivity of the 2D pattern typically changes from isolated discs to connected areas, such that the Euler value may become negative \cite{Mantz2008}.
Normalization of $r_\mathrm{D}$ using the square-root of the packing fraction, $\sqrt{\Phi}$, and the cell radius $R$ allows us to compare different cell densities.
We find that for small packing fractions, $U$, $F$, and $\chi$ show a distinctive different behavior compared to a random distribution (see Fig.~S3 in \footnotemark[2]). However, this is not the case for large $\Phi$ (see Fig.~S4 in \footnotemark[2]), where only $\chi$ is robust to morphological changes, thus using it as our main tool for comparison.

This procedure is applied to the momentary cell positions at any time, for which representative curves are shown as solid blue lines in Fig.~\ref{fig:FIG3}(b,c) for packing fractions of 12\% and 40\%, respectively.
At all times, the experimentally obtained cell positions are at variance with a distribution of randomly placed particles of the same density (dashed lines).
Notably, we find that for low cell densities (e.g.\ Fig.~\ref{fig:FIG3}(b)) the Euler number $\chi$ does not exhibit a time dependence: $\chi$ sharply decreases when $r_\mathrm{D}$ exceeds the cell radius $R$ and exhibits a minimum that is less pronounced and at a larger radius as compared to the random distributed particles.
While the Euler parameter does not change over time for cell densities up to $\mathrm{\Phi=32\%}$, we see a time dependency for filling fractions of $\mathrm{\Phi=38\%}$ and higher, see inset of Fig.~\ref{fig:FIG3}(c): the minimum of $\chi$ becomes successively more pronounced as time proceeds.
In general, as $r_\mathrm{D}$ increases, the cell positions overlap producing connected regions that reduce $\chi$. 
The disc radius $r_\mathrm{D}$ corresponding to the minimum in $\chi$ indicates the typical intercellular distance.
A faster decrease of $\chi$ compared to a random distribution means that there are significantly more cells with a distance $r_\mathrm{D}$ where this sudden decrease occurs, and is therefore indicative of the existence of clusters.
For $\Phi\leq 32\%$, the behavior of $\chi$ indicates local clusters, since the initial drop of is due to the mean intercellular distance within the cluster, while the secondary decrease to the shallow minimum is due to the mean intercluster distance, see Fig. \ref{fig:FIG3}(b).
However for $\Phi\geq 38\%$, there is a fully interconnected network, since the sudden decrease and the minimum of $\chi$ occur at the same $r_\mathrm{D}$, see Fig. \ref{fig:FIG3}(c).
The Euler number for a range of filling fractions is shown in Fig.~S5 in \footnotemark[2].

\textit{-- Simulations} To model the dynamics of the system several considerations have to be made. Figure~\ref{fig:FIG2}(a) shows that the movement of the cells is essentially confined to the direction of the flagella $\mathbf{\hat e_\phi }$, which changes over time according to some weak rotational diffusion. Furthermore, from the change in MSV we can discern that the average kinetic energy of the system decreases exponentially over time.
To capture these dynamics we propose the equation of motion based on a purely \emph{mechanistic} model
\begin{equation} \label{eq:2}
m \mathbf{\ddot{r}}_i = -\gamma \mathbf v_i + \theta \mathbf F_{\mathrm{tug}, i} + \sum_j\mathbf{h}_{ij}, 
%\mathbf F_i = -\gamma \mathbf v_i + \theta \mathbf F_\mathrm{tug, i} + \sum_j\mathbf{h}(\mathbf{r}_i-\mathbf{r}_j), 
\end{equation}
where $\mathbf{r}_i$ is the position of cell $i$,  $\mathbf v_i$ its velocity;
$ -\gamma \mathbf v_i$ represents viscous damping of the liquid medium, $\mathbf F_{\mathrm{tug}, i}$ the force generated by the flagella modulated by a coefficient $\theta\equiv\theta(t)=\theta_0e^{-t/t_0}+ \theta_1$, which decreases exponentially over time, and $\mathbf{h}_{ij}=k(r_{ij}-R_i-R_j)\mathbf{\hat{r}}_{ij}$ if $r_{ij}\le R_i+R_j$, $\mathbf{h}_{ij}=0$ otherwise, the purely repulsive harmonic interactions among cells $i$, $j$ at positions $\mathbf{r}_i$, $\mathbf{r}_j$, respectively, $r_{ij}=|\mathbf r_{ij}|=|\mathbf{r}_i-\mathbf{r}_j|$, $\mathbf{\hat{r}}_{ij}=\mathbf r_{ij}/r_{ij}$.  The coefficients $\theta_0$ and $\theta_1$ were chosen so that the average kinetic energy is the same as in the experiments. The magnitude of the force on each flagellum, $F_{1,2}$, is independently drawn from an exponential distribution, and their direction is parallel or antiparallel to the flagellar orientation; thus the total flagellar tug reads
$ \mathbf F_{\mathrm{tug}, i} = (F_1 - F_2) ~ \mathbf{\hat e_\phi}$. We integrate Newton's equations for a polydisperse population of cells \footnotemark[2] where each cell is subject to the force in Eq.~\eqref{eq:2} using standard methods \cite{allen2017}.
As shown in the inset of Fig.~\ref{fig:FIG3}(b), this approach cannot capture the final cell configuration, see inset of Fig.~\ref{fig:FIG3}(b).

In order to capture the experimental data, we extend our simulation approach towards implementing a mechanosensing mechanism \cite{fujiu2011mechanoreception,collingridge2013compartmentalized}, which we term cognitive cell-cell interactions. 
Extending a recent approach~\cite{hornischer2019structural,hornischer2022,wissnerPRL2013}, we define a cognitive force associated to the cell's mechanosensing and exploration of its surrounding
\begin{equation} \label{eq:4}
  \mathbf{F}_c \propto \left< \sum_{n=1}^{N_\Omega} \mathbf{f}_n(0) \ln \left(\frac{\Omega_n}{\left<\Omega_n\right>}\right) \right>\,.
\end{equation}
The computation of $\mathbf{F}_c$ relies on the calculation of a Boltzmann--Shannon entropy measuring the information content of the surrounding environment, specifically, the location of the neighboring cells. 
This information is sampled statistically with $N_\Omega$ sampling Brownian trajectories emanating from the current cell position and interacting (only via $\mathbf{h}$) with the other cells. The sampling trajectories collectively form the `cognitive map' of each cell, whose average size is equal to $11.9$ \textmu m, which corresponds to the average length of the flagella. 
For details, see \footnotemark[2].
The force $\mathbf{F}_c$ in a given direction $\mathbf{f}_n(0)$ is weighted by the space available to the sampling trajectories starting in the same direction $\mathbf{f}_n(0)$.
%
%the first step of the sampling trajectory $\mathbf{f}_n(0)$ and $\Omega_n$ as the information entropy associated to a single trajectory, which will be maximised at the time horizon defined by the duration of the trajectories.
%
As the \textit{Chlamydomonas} cells exhibit clustering dynamics, we propose that $\Omega_n$ has to minimise the available space around the cells. The radius of gyration is a natural choice, and we arrive at $\Omega_n = \mathcal{R}^2_{max} - \mathcal{R}^2_n$,
where $\mathcal{R}^2_n$ is the square radius of gyration associated to the trajectory $n$ and $\mathcal{R}^2_{max}$ the maximum square radius of gyration for all $N_\Omega$ trajectories.
Finally, the cognitive force is projected onto the direction of the flagella $\mathbf{\hat e_\phi}$, $\mathbf F^\mathrm{fl}_{\mathrm{c}, i} = (\mathbf{F}_{\mathrm{c}, i} \cdot \mathbf{\hat e}_\phi ) \mathbf{\hat e}_\phi$,
which replaces $\mathbf F_{\mathrm{tug}, i}$ in Eq.~\eqref{eq:2}. The equation of motion for each cell in the \emph{cognitive} model reads as
\begin{equation} \label{eq:cog_mod}
m \mathbf{\ddot{r}}_i  = -\gamma \mathbf v_i + \theta_c \mathbf F^\mathrm{fl}_{\mathrm{c}, i} + \sum_j\mathbf{h}_{ij}\,.
%\mathbf F = -\gamma \mathbf v + \theta_c \mathbf F^{fl}_c + \mathbf{h}(\mathbf{r}_i-\mathbf{r}_j)\,.
\end{equation}
Following the observed intermittency of the cells (see Fig.~\ref{fig:FIG2}), we apply the cognitive force $\mathbf{F}_c$ intermittently. At each time step we draw a uniformly distributed random number $r\in [0,1]$ such that $\theta_c=1$ if $r<\left[(v^2_{max}-v^2_{min})\exp{(-t/t_0)}+v^2_{min}\right]/v^2_{max}$, and $\theta_c=0$ otherwise.

Newton's equations with the force in Eq.~\eqref{eq:cog_mod} for all cells at any timestep are integrated using standard molecular dynamics methods \cite{allen2017}.
We find that the cognitive model can reproduce both the MSV (Fig.~\ref{fig:FIG2}(b)) and the morphological features as measured by the three Minkowski functional (Fig.~\ref{fig:FIG3}), which was confirmed with a residual analysis (see Fig.~S6 in \footnotemark[2]).

\begin{figure}
    \centering
    \includegraphics{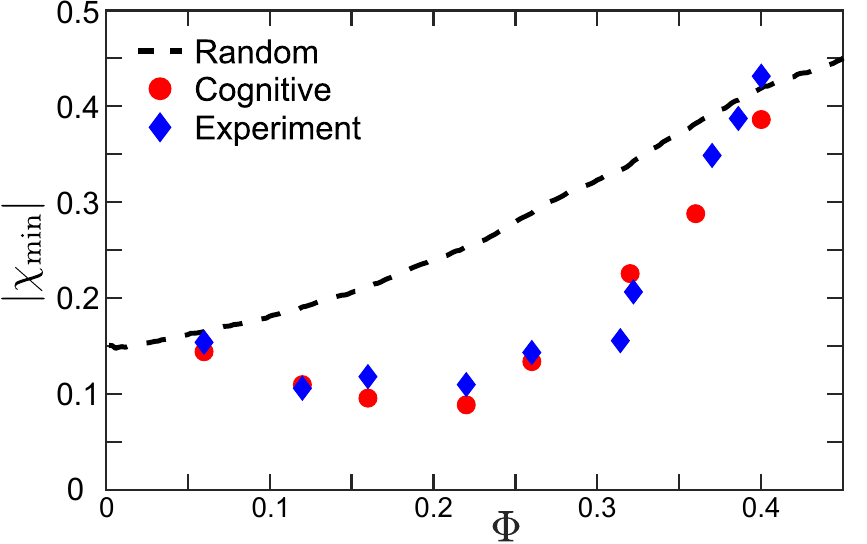}
    \caption{Absolute value of the minima of the Euler number $|\chi_\mathrm{min}|$ at late times in the experiments (blue diamonds), simulations of the cognitive model (red circles), and for randomly distributed particles (black dashed line). The error of $|\chi_\mathrm{min}|$ are within the symbol, and thus not shown.
    }
    \label{fig:FIG4}
\end{figure}

\textit{-- Discussion}
Overall, we use the depth of the minimum of the Euler number, $|\chi_\mathrm{min}|$, for quantitative comparison of experimental data, randomly distributed objects and theoretical modelling, which is displayed in Fig.~\ref{fig:FIG4}.
For less dense populations, $|\chi_\mathrm{min}|$ in experiments is found to be smaller than for randomly distributed particles (with and without mechanistic activity).
At a packing fraction $\Phi\approx 34\%$, we see a steep increase of the minimum, which eventually crosses the random model prediction.
This significant change in $|\chi_\mathrm{min}|$ is accompanied by mesoscopic changes of the cell distribution, where the local cell clusters become interconnected for packing fractions $\Phi>$34\% (see Fig.~S7 in \footnotemark[2]).

We can describe this change in global connectivity using percolation theory.
Spherical 2D particles typically form triangular lattices, which has a site percolation threshold of 0.5 \cite{van1997percolation}.
Using the random close packing of 2D spheres $\Phi_\mathrm{RCP}=82\%$ \cite{meyer2010jamming}, we estimate the percolation threshold for our system to be $\Phi_\mathrm{per}=41\%$.
The difference with the observed threshold can be attributed to the tendency of the cells to cluster and the fact that our system does not form exact triangular lattices.
This percolation threshold is quantitatively captured by our model that incorporates cognitive forces (see Fig.~S8 in \footnotemark[2]).

%%%%%%%%%%%%%%%%%%%%%
%%%%%%%%%%%%%%%%%%%%%

The need to include cognitive forces in the model to reproduce the clustering observed in the experiments indicates that there is an active mechanism that drives the cells together.
As aforementioned, a rapid cell motion is attributed to an elevated Ca$^{2+}$ signal in a flagellum. However, it has been shown that a Ca$^{2+}$ signal can be induced by exerting a mechanical stress on the flagellum \cite{collingridge2013compartmentalized}.
A flagellum that has less available area around it, due to the presence of other cells, will have fewer adhesion sites with the substrate and thus experience smaller mechanical stresses.
However, a flagellum away from clusters will have a larger available area and thus more adhesion sites, leading to a higher mechanical stress exerted on the flagellum.
As a result, it will experience more frequent elevations of Ca$^{2+}$ signals causing it to move statistically away from diluted regions, thus promoting cluster formation.
Our cognitive force offers an economic model of the sensory processes underpinned by the Ca$^{2+}$-dynamics in the flagella, by identifying the size of the cognitive map with the flagellar size.

\textit{-- Conclusion} In conclusion, we employed the Euler number, one of the three Minkowski functionals for 2D systems, to dissect the role of gliding motility with regard to the self-organization and clustering of surface-associated \textit{Chlamydomonas} cells. 
A cognitive-force model representing the flagellar exploration of available space and its mechanosensory response can, in contrast to classical mechanistic approaches, reproduce the non-random cell positions obtained in the experiments.

We find that the gliding motility is a key mechanism for the formation of a compact monolayer of \textit{Chlamydomonas} cells on a surface, which represents a favourable configuration for these photoactive microorganisms to perform photosynthesis. 
Swimming motility assisted by phototaxis is essential for \textit{Chlamydomonas} in their natural habitats to map their environment for light sources.
Once the cells have found optimal light conditions for photosynthesis, their gliding motility mode enables the population to form compact surface-bound monolayers for highly efficient light harvesting.

The authors thank the G\"ottingen Algae Culture Collection (SAG) for providing the \textit{Chlamydomonas reinhardtii} wild-type strain SAG~11-32b. We acknowledge S.~Klumpp, S.~Herminghaus and M.~Lorenz for fruitful discussions.


\begin{thebibliography}{43}%
\makeatletter
\providecommand \@ifxundefined [1]{%
 \@ifx{#1\undefined}
}%
\providecommand \@ifnum [1]{%
 \ifnum #1\expandafter \@firstoftwo
 \else \expandafter \@secondoftwo
 \fi
}%
\providecommand \@ifx [1]{%
 \ifx #1\expandafter \@firstoftwo
 \else \expandafter \@secondoftwo
 \fi
}%
\providecommand \natexlab [1]{#1}%
\providecommand \enquote  [1]{``#1''}%
\providecommand \bibnamefont  [1]{#1}%
\providecommand \bibfnamefont [1]{#1}%
\providecommand \citenamefont [1]{#1}%
\providecommand \href@noop [0]{\@secondoftwo}%
\providecommand \href [0]{\begingroup \@sanitize@url \@href}%
\providecommand \@href[1]{\@@startlink{#1}\@@href}%
\providecommand \@@href[1]{\endgroup#1\@@endlink}%
\providecommand \@sanitize@url [0]{\catcode `\\12\catcode `\$12\catcode
  `\&12\catcode `\#12\catcode `\^12\catcode `\_12\catcode `\%12\relax}%
\providecommand \@@startlink[1]{}%
\providecommand \@@endlink[0]{}%
\providecommand \url  [0]{\begingroup\@sanitize@url \@url }%
\providecommand \@url [1]{\endgroup\@href {#1}{\urlprefix }}%
\providecommand \urlprefix  [0]{URL }%
\providecommand \Eprint [0]{\href }%
\providecommand \doibase [0]{https://doi.org/}%
\providecommand \selectlanguage [0]{\@gobble}%
\providecommand \bibinfo  [0]{\@secondoftwo}%
\providecommand \bibfield  [0]{\@secondoftwo}%
\providecommand \translation [1]{[#1]}%
\providecommand \BibitemOpen [0]{}%
\providecommand \bibitemStop [0]{}%
\providecommand \bibitemNoStop [0]{.\EOS\space}%
\providecommand \EOS [0]{\spacefactor3000\relax}%
\providecommand \BibitemShut  [1]{\csname bibitem#1\endcsname}%
\let\auto@bib@innerbib\@empty
%</preamble>
\bibitem [{\citenamefont {Berg}\ and\ \citenamefont {Brown}(1972)}]{Berg1973}%
  \BibitemOpen
  \bibfield  {author} {\bibinfo {author} {\bibfnamefont {H.~C.}\ \bibnamefont
  {Berg}}\ and\ \bibinfo {author} {\bibfnamefont {D.~A.}\ \bibnamefont
  {Brown}},\ }\bibfield  {title} {\bibinfo {title} {Chemotaxis in {E}scherichia
  coli analysed by three-dimensional tracking},\ }\href@noop {} {\bibfield
  {journal} {\bibinfo  {journal} {Nature}\ }\textbf {\bibinfo {volume} {239}},\
  \bibinfo {pages} {500} (\bibinfo {year} {1972})}\BibitemShut {NoStop}%
\bibitem [{\citenamefont {Miller}\ and\ \citenamefont
  {Bassler}(2001)}]{Miller2001}%
  \BibitemOpen
  \bibfield  {author} {\bibinfo {author} {\bibfnamefont {M.~B.}\ \bibnamefont
  {Miller}}\ and\ \bibinfo {author} {\bibfnamefont {B.~L.}\ \bibnamefont
  {Bassler}},\ }\bibfield  {title} {\bibinfo {title} {Quorum sensing in
  bacteria},\ }\href {https://doi.org/10.1146/annurev.micro.55.1.165}
  {\bibfield  {journal} {\bibinfo  {journal} {Annu. Rev. Microbiol.}\ }\textbf
  {\bibinfo {volume} {55}},\ \bibinfo {pages} {165} (\bibinfo {year}
  {2001})}\BibitemShut {NoStop}%
\bibitem [{\citenamefont {Lauga}\ and\ \citenamefont
  {Powers}(2009)}]{Lauga_2009}%
  \BibitemOpen
  \bibfield  {author} {\bibinfo {author} {\bibfnamefont {E.}~\bibnamefont
  {Lauga}}\ and\ \bibinfo {author} {\bibfnamefont {T.~R.}\ \bibnamefont
  {Powers}},\ }\bibfield  {title} {\bibinfo {title} {The hydrodynamics of
  swimming microorganisms},\ }\href
  {https://doi.org/10.1088/0034-4885/72/9/096601} {\bibfield  {journal}
  {\bibinfo  {journal} {Rep. Prog. Phys.}\ }\textbf {\bibinfo {volume} {72}},\
  \bibinfo {pages} {096601} (\bibinfo {year} {2009})}\BibitemShut {NoStop}%
\bibitem [{\citenamefont {Schulman}\ \emph {et~al.}(2014)\citenamefont
  {Schulman}, \citenamefont {Backholm}, \citenamefont {Ryu},\ and\
  \citenamefont {Dalnoki-Veress}}]{Schulman2014}%
  \BibitemOpen
  \bibfield  {author} {\bibinfo {author} {\bibfnamefont {R.~D.}\ \bibnamefont
  {Schulman}}, \bibinfo {author} {\bibfnamefont {M.}~\bibnamefont {Backholm}},
  \bibinfo {author} {\bibfnamefont {W.~S.}\ \bibnamefont {Ryu}},\ and\ \bibinfo
  {author} {\bibfnamefont {K.}~\bibnamefont {Dalnoki-Veress}},\ }\bibfield
  {title} {\bibinfo {title} {Dynamic force patterns of an undulatory
  microswimmer},\ }\href {https://doi.org/10.1103/PhysRevE.89.050701}
  {\bibfield  {journal} {\bibinfo  {journal} {Phys. Rev. E}\ }\textbf {\bibinfo
  {volume} {89}},\ \bibinfo {pages} {050701} (\bibinfo {year}
  {2014})}\BibitemShut {NoStop}%
\bibitem [{\citenamefont {Noselli}\ \emph {et~al.}(2019)\citenamefont
  {Noselli}, \citenamefont {Beran}, \citenamefont {Arroyo},\ and\ \citenamefont
  {DeSimone}}]{Noselli2019}%
  \BibitemOpen
  \bibfield  {author} {\bibinfo {author} {\bibfnamefont {G.}~\bibnamefont
  {Noselli}}, \bibinfo {author} {\bibfnamefont {A.}~\bibnamefont {Beran}},
  \bibinfo {author} {\bibfnamefont {M.}~\bibnamefont {Arroyo}},\ and\ \bibinfo
  {author} {\bibfnamefont {A.}~\bibnamefont {DeSimone}},\ }\bibfield  {title}
  {\bibinfo {title} {Swimming euglena respond to confinement with a behavioural
  change enabling effective crawling},\ }\href@noop {} {\bibfield  {journal}
  {\bibinfo  {journal} {Nat. Phys.}\ }\textbf {\bibinfo {volume} {15}},\
  \bibinfo {pages} {496–502} (\bibinfo {year} {2019})}\BibitemShut {NoStop}%
\bibitem [{\citenamefont {Mitchinson}\ and\ \citenamefont
  {Cramer}(1996)}]{Mitchinson1996}%
  \BibitemOpen
  \bibfield  {author} {\bibinfo {author} {\bibfnamefont {T.}~\bibnamefont
  {Mitchinson}}\ and\ \bibinfo {author} {\bibfnamefont {L.}~\bibnamefont
  {Cramer}},\ }\bibfield  {title} {\bibinfo {title} {Actin-based cell motility
  and cell locomotion},\ }\href@noop {} {\bibfield  {journal} {\bibinfo
  {journal} {Cell}\ }\textbf {\bibinfo {volume} {84}},\ \bibinfo {pages} {371}
  (\bibinfo {year} {1996})}\BibitemShut {NoStop}%
\bibitem [{\citenamefont {Batchelder}\ \emph {et~al.}(2011)\citenamefont
  {Batchelder}, \citenamefont {Hollopeter}, \citenamefont {Campillo},
  \citenamefont {Mezanges}, \citenamefont {Jorgensen}, \citenamefont {Nassoy},
  \citenamefont {Sens},\ and\ \citenamefont {Plastino}}]{Batchelder2011}%
  \BibitemOpen
  \bibfield  {author} {\bibinfo {author} {\bibfnamefont {E.~L.}\ \bibnamefont
  {Batchelder}}, \bibinfo {author} {\bibfnamefont {G.}~\bibnamefont
  {Hollopeter}}, \bibinfo {author} {\bibfnamefont {C.}~\bibnamefont
  {Campillo}}, \bibinfo {author} {\bibfnamefont {X.}~\bibnamefont {Mezanges}},
  \bibinfo {author} {\bibfnamefont {E.~M.}\ \bibnamefont {Jorgensen}}, \bibinfo
  {author} {\bibfnamefont {P.}~\bibnamefont {Nassoy}}, \bibinfo {author}
  {\bibfnamefont {P.}~\bibnamefont {Sens}},\ and\ \bibinfo {author}
  {\bibfnamefont {J.}~\bibnamefont {Plastino}},\ }\bibfield  {title} {\bibinfo
  {title} {Membrane tension regulates motility by controlling lamellipodium
  organization},\ }\href {https://doi.org/10.1073/pnas.1010481108} {\bibfield
  {journal} {\bibinfo  {journal} {Proc. Natl. Acad. Sci. USA}\ }\textbf
  {\bibinfo {volume} {108}},\ \bibinfo {pages} {11429} (\bibinfo {year}
  {2011})}\BibitemShut {NoStop}%
\bibitem [{\citenamefont {Rossi}\ \emph {et~al.}(2017)\citenamefont {Rossi},
  \citenamefont {Cicconofri}, \citenamefont {Beran}, \citenamefont {Noselli},\
  and\ \citenamefont {DeSimone}}]{Rossi13085}%
  \BibitemOpen
  \bibfield  {author} {\bibinfo {author} {\bibfnamefont {M.}~\bibnamefont
  {Rossi}}, \bibinfo {author} {\bibfnamefont {G.}~\bibnamefont {Cicconofri}},
  \bibinfo {author} {\bibfnamefont {A.}~\bibnamefont {Beran}}, \bibinfo
  {author} {\bibfnamefont {G.}~\bibnamefont {Noselli}},\ and\ \bibinfo {author}
  {\bibfnamefont {A.}~\bibnamefont {DeSimone}},\ }\bibfield  {title} {\bibinfo
  {title} {Kinematics of flagellar swimming in {E}uglena gracilis: {H}elical
  trajectories and flagellar shapes},\ }\href
  {https://doi.org/10.1073/pnas.1708064114} {\bibfield  {journal} {\bibinfo
  {journal} {Proc. Natl. Acad. Sci. USA}\ }\textbf {\bibinfo {volume} {114}},\
  \bibinfo {pages} {13085} (\bibinfo {year} {2017})}\BibitemShut {NoStop}%
\bibitem [{\citenamefont {Wan}\ and\ \citenamefont
  {Goldstein}(2018)}]{Wan2018}%
  \BibitemOpen
  \bibfield  {author} {\bibinfo {author} {\bibfnamefont {K.~Y.}\ \bibnamefont
  {Wan}}\ and\ \bibinfo {author} {\bibfnamefont {R.~E.}\ \bibnamefont
  {Goldstein}},\ }\bibfield  {title} {\bibinfo {title} {Time irreversibility
  and criticality in the motility of a flagellate microorganism},\ }\href
  {https://doi.org/10.1103/PhysRevLett.121.058103} {\bibfield  {journal}
  {\bibinfo  {journal} {Phys. Rev. Lett.}\ }\textbf {\bibinfo {volume} {121}},\
  \bibinfo {pages} {058103} (\bibinfo {year} {2018})}\BibitemShut {NoStop}%
\bibitem [{\citenamefont {Tsang}\ \emph {et~al.}(2018)\citenamefont {Tsang},
  \citenamefont {Lam},\ and\ \citenamefont {Riedel-Kruse}}]{Tsang2018}%
  \BibitemOpen
  \bibfield  {author} {\bibinfo {author} {\bibfnamefont {A.~C.~H.}\
  \bibnamefont {Tsang}}, \bibinfo {author} {\bibfnamefont {A.~T.}\ \bibnamefont
  {Lam}},\ and\ \bibinfo {author} {\bibfnamefont {I.~H.}\ \bibnamefont
  {Riedel-Kruse}},\ }\bibfield  {title} {\bibinfo {title} {Polygonal motion and
  adaptable phototaxis via flagellar beat switching in the microswimmer euglena
  gracilis},\ }\href@noop {} {\bibfield  {journal} {\bibinfo  {journal} {Nat.
  Phys.}\ }\textbf {\bibinfo {volume} {14}},\ \bibinfo {pages} {1216–1222}
  (\bibinfo {year} {2018})}\BibitemShut {NoStop}%
\bibitem [{\citenamefont {Rüffer}\ and\ \citenamefont
  {Nultsch}(1985)}]{Ruffer1985}%
  \BibitemOpen
  \bibfield  {author} {\bibinfo {author} {\bibfnamefont {U.}~\bibnamefont
  {Rüffer}}\ and\ \bibinfo {author} {\bibfnamefont {W.}~\bibnamefont
  {Nultsch}},\ }\bibfield  {title} {\bibinfo {title} {High-speed
  cinematographic analysis of the movement of chlamydomonas},\ }\href
  {https://doi.org/https://doi.org/10.1002/cm.970050307} {\bibfield  {journal}
  {\bibinfo  {journal} {Cell Motil.}\ }\textbf {\bibinfo {volume} {5}},\
  \bibinfo {pages} {251} (\bibinfo {year} {1985})}\BibitemShut {NoStop}%
\bibitem [{\citenamefont {Drescher}\ \emph {et~al.}(2010)\citenamefont
  {Drescher}, \citenamefont {Goldstein}, \citenamefont {Michel}, \citenamefont
  {Polin},\ and\ \citenamefont {Tuval}}]{drescher2010}%
  \BibitemOpen
  \bibfield  {author} {\bibinfo {author} {\bibfnamefont {K.}~\bibnamefont
  {Drescher}}, \bibinfo {author} {\bibfnamefont {R.~E.}\ \bibnamefont
  {Goldstein}}, \bibinfo {author} {\bibfnamefont {N.}~\bibnamefont {Michel}},
  \bibinfo {author} {\bibfnamefont {M.}~\bibnamefont {Polin}},\ and\ \bibinfo
  {author} {\bibfnamefont {I.}~\bibnamefont {Tuval}},\ }\bibfield  {title}
  {\bibinfo {title} {Direct measurement of the flow field around swimming
  microorganisms},\ }\href {https://doi.org/10.1103/PhysRevLett.105.168101}
  {\bibfield  {journal} {\bibinfo  {journal} {Phys. Rev. Lett.}\ }\textbf
  {\bibinfo {volume} {105}},\ \bibinfo {pages} {168101} (\bibinfo {year}
  {2010})}\BibitemShut {NoStop}%
\bibitem [{\citenamefont {Friedrich}\ and\ \citenamefont
  {J\"ulicher}(2012)}]{friedrich2012}%
  \BibitemOpen
  \bibfield  {author} {\bibinfo {author} {\bibfnamefont {B.~M.}\ \bibnamefont
  {Friedrich}}\ and\ \bibinfo {author} {\bibfnamefont {F.}~\bibnamefont
  {J\"ulicher}},\ }\bibfield  {title} {\bibinfo {title} {Flagellar
  synchronization independent of hydrodynamic interactions},\ }\href
  {https://doi.org/10.1103/PhysRevLett.109.138102} {\bibfield  {journal}
  {\bibinfo  {journal} {Phys. Rev. Lett.}\ }\textbf {\bibinfo {volume} {109}},\
  \bibinfo {pages} {138102} (\bibinfo {year} {2012})}\BibitemShut {NoStop}%
\bibitem [{\citenamefont {Geyer}\ \emph {et~al.}(2013)\citenamefont {Geyer},
  \citenamefont {J{\"u}licher}, \citenamefont {Howard},\ and\ \citenamefont
  {Friedrich}}]{geyer2013}%
  \BibitemOpen
  \bibfield  {author} {\bibinfo {author} {\bibfnamefont {V.~F.}\ \bibnamefont
  {Geyer}}, \bibinfo {author} {\bibfnamefont {F.}~\bibnamefont {J{\"u}licher}},
  \bibinfo {author} {\bibfnamefont {J.}~\bibnamefont {Howard}},\ and\ \bibinfo
  {author} {\bibfnamefont {B.~M.}\ \bibnamefont {Friedrich}},\ }\bibfield
  {title} {\bibinfo {title} {Cell-body rocking is a dominant mechanism for
  flagellar synchronization in a swimming alga},\ }\href@noop {} {\bibfield
  {journal} {\bibinfo  {journal} {Proc. Natl. Acad. Sci. USA}\ }\textbf
  {\bibinfo {volume} {110}},\ \bibinfo {pages} {18058} (\bibinfo {year}
  {2013})}\BibitemShut {NoStop}%
\bibitem [{\citenamefont {Wan}\ and\ \citenamefont
  {Goldstein}(2016)}]{wan2016coordinated}%
  \BibitemOpen
  \bibfield  {author} {\bibinfo {author} {\bibfnamefont {K.~Y.}\ \bibnamefont
  {Wan}}\ and\ \bibinfo {author} {\bibfnamefont {R.~E.}\ \bibnamefont
  {Goldstein}},\ }\bibfield  {title} {\bibinfo {title} {Coordinated beating of
  algal flagella is mediated by basal coupling},\ }\href@noop {} {\bibfield
  {journal} {\bibinfo  {journal} {Proc. Natl. Acad. Sci. USA}\ }\textbf
  {\bibinfo {volume} {113}},\ \bibinfo {pages} {E2784} (\bibinfo {year}
  {2016})}\BibitemShut {NoStop}%
\bibitem [{\citenamefont {B\"oddeker}\ \emph {et~al.}(2020)\citenamefont
  {B\"oddeker}, \citenamefont {Karpitschka}, \citenamefont {Kreis},
  \citenamefont {Magdelaine},\ and\ \citenamefont
  {B\"aumchen}}]{Boeddeker2020}%
  \BibitemOpen
  \bibfield  {author} {\bibinfo {author} {\bibfnamefont {T.~J.}\ \bibnamefont
  {B\"oddeker}}, \bibinfo {author} {\bibfnamefont {S.}~\bibnamefont
  {Karpitschka}}, \bibinfo {author} {\bibfnamefont {C.~T.}\ \bibnamefont
  {Kreis}}, \bibinfo {author} {\bibfnamefont {Q.}~\bibnamefont {Magdelaine}},\
  and\ \bibinfo {author} {\bibfnamefont {O.}~\bibnamefont {B\"aumchen}},\
  }\bibfield  {title} {\bibinfo {title} {Dynamic force measurements on swimming
  {C}hlamydomonas cells using micropipette force sensors},\ }\href
  {https://doi.org/10.1098/rsif.2019.0580} {\bibfield  {journal} {\bibinfo
  {journal} {J. R. Soc. Interface}\ }\textbf {\bibinfo {volume} {17}},\
  \bibinfo {pages} {20190580} (\bibinfo {year} {2020})}\BibitemShut {NoStop}%
\bibitem [{\citenamefont {Polin}\ \emph {et~al.}(2009)\citenamefont {Polin},
  \citenamefont {Tuval}, \citenamefont {Drescher}, \citenamefont {Gollub},\
  and\ \citenamefont {Goldstein}}]{polin2009}%
  \BibitemOpen
  \bibfield  {author} {\bibinfo {author} {\bibfnamefont {M.}~\bibnamefont
  {Polin}}, \bibinfo {author} {\bibfnamefont {I.}~\bibnamefont {Tuval}},
  \bibinfo {author} {\bibfnamefont {K.}~\bibnamefont {Drescher}}, \bibinfo
  {author} {\bibfnamefont {J.~P.}\ \bibnamefont {Gollub}},\ and\ \bibinfo
  {author} {\bibfnamefont {R.~E.}\ \bibnamefont {Goldstein}},\ }\bibfield
  {title} {\bibinfo {title} {Chlamydomonas swims with two “gears” in a
  eukaryotic version of run-and-tumble locomotion},\ }\href@noop {} {\bibfield
  {journal} {\bibinfo  {journal} {Science}\ }\textbf {\bibinfo {volume}
  {325}},\ \bibinfo {pages} {487} (\bibinfo {year} {2009})}\BibitemShut
  {NoStop}%
\bibitem [{\citenamefont {Ostapenko}\ \emph {et~al.}(2018)\citenamefont
  {Ostapenko}, \citenamefont {Schwarzendahl}, \citenamefont {B{\"o}ddeker},
  \citenamefont {Kreis}, \citenamefont {Cammann}, \citenamefont {Mazza},\ and\
  \citenamefont {B{\"a}umchen}}]{ostapenko2018curvature}%
  \BibitemOpen
  \bibfield  {author} {\bibinfo {author} {\bibfnamefont {T.}~\bibnamefont
  {Ostapenko}}, \bibinfo {author} {\bibfnamefont {F.~J.}\ \bibnamefont
  {Schwarzendahl}}, \bibinfo {author} {\bibfnamefont {T.~J.}\ \bibnamefont
  {B{\"o}ddeker}}, \bibinfo {author} {\bibfnamefont {C.~T.}\ \bibnamefont
  {Kreis}}, \bibinfo {author} {\bibfnamefont {J.}~\bibnamefont {Cammann}},
  \bibinfo {author} {\bibfnamefont {M.~G.}\ \bibnamefont {Mazza}},\ and\
  \bibinfo {author} {\bibfnamefont {O.}~\bibnamefont {B{\"a}umchen}},\
  }\bibfield  {title} {\bibinfo {title} {Curvature-guided motility of
  microalgae in geometric confinement},\ }\href@noop {} {\bibfield  {journal}
  {\bibinfo  {journal} {Phys. Rev. Lett.}\ }\textbf {\bibinfo {volume} {120}},\
  \bibinfo {pages} {068002} (\bibinfo {year} {2018})}\BibitemShut {NoStop}%
\bibitem [{\citenamefont {Harris}\ \emph {et~al.}(2009)\citenamefont {Harris},
  \citenamefont {Stern},\ and\ \citenamefont
  {Witman}}]{harris2009chlamydomonas}%
  \BibitemOpen
  \bibfield  {author} {\bibinfo {author} {\bibfnamefont {E.~H.}\ \bibnamefont
  {Harris}}, \bibinfo {author} {\bibfnamefont {D.~B.}\ \bibnamefont {Stern}},\
  and\ \bibinfo {author} {\bibfnamefont {G.~B.}\ \bibnamefont {Witman}},\
  }\href@noop {} {\emph {\bibinfo {title} {The Chlamydomonas Sourcebook}}},\
  Vol.~\bibinfo {volume} {1}\ (\bibinfo  {publisher} {Elsevier, San Diego,
  CA},\ \bibinfo {year} {2009})\BibitemShut {NoStop}%
\bibitem [{\citenamefont {Kreis}\ \emph {et~al.}(2018)\citenamefont {Kreis},
  \citenamefont {Le~Blay}, \citenamefont {Linne}, \citenamefont {Makowski},\
  and\ \citenamefont {B{\"a}umchen}}]{kreis2018adhesion}%
  \BibitemOpen
  \bibfield  {author} {\bibinfo {author} {\bibfnamefont {C.~T.}\ \bibnamefont
  {Kreis}}, \bibinfo {author} {\bibfnamefont {M.}~\bibnamefont {Le~Blay}},
  \bibinfo {author} {\bibfnamefont {C.}~\bibnamefont {Linne}}, \bibinfo
  {author} {\bibfnamefont {M.~M.}\ \bibnamefont {Makowski}},\ and\ \bibinfo
  {author} {\bibfnamefont {O.}~\bibnamefont {B{\"a}umchen}},\ }\bibfield
  {title} {\bibinfo {title} {Adhesion of chlamydomonas microalgae to surfaces
  is switchable by light},\ }\href@noop {} {\bibfield  {journal} {\bibinfo
  {journal} {Nat. Phys.}\ }\textbf {\bibinfo {volume} {14}},\ \bibinfo {pages}
  {45} (\bibinfo {year} {2018})}\BibitemShut {NoStop}%
\bibitem [{\citenamefont {Kreis}\ \emph {et~al.}(2019)\citenamefont {Kreis},
  \citenamefont {Grangier},\ and\ \citenamefont {Bäumchen}}]{Kreis2019}%
  \BibitemOpen
  \bibfield  {author} {\bibinfo {author} {\bibfnamefont {C.~T.}\ \bibnamefont
  {Kreis}}, \bibinfo {author} {\bibfnamefont {A.}~\bibnamefont {Grangier}},\
  and\ \bibinfo {author} {\bibfnamefont {O.}~\bibnamefont {Bäumchen}},\
  }\bibfield  {title} {\bibinfo {title} {In vivo adhesion force measurements of
  {C}hlamydomonas on model substrates},\ }\href
  {https://doi.org/10.1039/C8SM02236D} {\bibfield  {journal} {\bibinfo
  {journal} {Soft Matter}\ }\textbf {\bibinfo {volume} {15}},\ \bibinfo {pages}
  {3027} (\bibinfo {year} {2019})}\BibitemShut {NoStop}%
\bibitem [{\citenamefont {Stepanek}\ and\ \citenamefont
  {Pigino}(2016)}]{Stepanek2016}%
  \BibitemOpen
  \bibfield  {author} {\bibinfo {author} {\bibfnamefont {L.}~\bibnamefont
  {Stepanek}}\ and\ \bibinfo {author} {\bibfnamefont {G.}~\bibnamefont
  {Pigino}},\ }\bibfield  {title} {\bibinfo {title} {Microtubule doublets are
  double-track railways for intraflagellar transport trains},\ }\href
  {https://doi.org/10.1126/science.aaf4594} {\bibfield  {journal} {\bibinfo
  {journal} {Science}\ }\textbf {\bibinfo {volume} {352}},\ \bibinfo {pages}
  {721} (\bibinfo {year} {2016})}\BibitemShut {NoStop}%
\bibitem [{\citenamefont {Shih}\ \emph {et~al.}(2013)\citenamefont {Shih},
  \citenamefont {Engel}, \citenamefont {Kocabas}, \citenamefont {Bilyard},
  \citenamefont {Gennerich}, \citenamefont {Marshall},\ and\ \citenamefont
  {Yildiz}}]{Shih2013}%
  \BibitemOpen
  \bibfield  {author} {\bibinfo {author} {\bibfnamefont {S.~M.}\ \bibnamefont
  {Shih}}, \bibinfo {author} {\bibfnamefont {B.~D.}\ \bibnamefont {Engel}},
  \bibinfo {author} {\bibfnamefont {F.}~\bibnamefont {Kocabas}}, \bibinfo
  {author} {\bibfnamefont {T.}~\bibnamefont {Bilyard}}, \bibinfo {author}
  {\bibfnamefont {A.}~\bibnamefont {Gennerich}}, \bibinfo {author}
  {\bibfnamefont {W.~F.}\ \bibnamefont {Marshall}},\ and\ \bibinfo {author}
  {\bibfnamefont {A.}~\bibnamefont {Yildiz}},\ }\bibfield  {title} {\bibinfo
  {title} {Intraflagellar transport drives flagellar surface motility},\ }\href
  {https://doi.org/10.7554/eLife.00744} {\bibfield  {journal} {\bibinfo
  {journal} {eLife}\ }\textbf {\bibinfo {volume} {2}},\ \bibinfo {pages}
  {e00744} (\bibinfo {year} {2013})}\BibitemShut {NoStop}%
\bibitem [{\citenamefont {Bloodgood}(1981)}]{Bloodgood1981}%
  \BibitemOpen
  \bibfield  {author} {\bibinfo {author} {\bibfnamefont {R.}~\bibnamefont
  {Bloodgood}},\ }\bibfield  {title} {\bibinfo {title} {Flagella-dependent
  gliding motility in chlamydomonas},\ }\href@noop {} {\bibfield  {journal}
  {\bibinfo  {journal} {Protoplasma}\ }\textbf {\bibinfo {volume} {106}},\
  \bibinfo {pages} {183–192} (\bibinfo {year} {1981})}\BibitemShut {NoStop}%
\bibitem [{\citenamefont {Xu}\ \emph {et~al.}(2020)\citenamefont {Xu},
  \citenamefont {Oltmanns}, \citenamefont {Zhao}, \citenamefont {Girot},
  \citenamefont {Karimi}, \citenamefont {Hoepfner}, \citenamefont {Kelterborn},
  \citenamefont {Scholz}, \citenamefont {Beißel}, \citenamefont {Hegemann},
  \citenamefont {Bäumchen}, \citenamefont {Liu}, \citenamefont {Huang},\ and\
  \citenamefont {Hippler}}]{Xu2020}%
  \BibitemOpen
  \bibfield  {author} {\bibinfo {author} {\bibfnamefont {N.}~\bibnamefont
  {Xu}}, \bibinfo {author} {\bibfnamefont {A.}~\bibnamefont {Oltmanns}},
  \bibinfo {author} {\bibfnamefont {L.}~\bibnamefont {Zhao}}, \bibinfo {author}
  {\bibfnamefont {A.}~\bibnamefont {Girot}}, \bibinfo {author} {\bibfnamefont
  {M.}~\bibnamefont {Karimi}}, \bibinfo {author} {\bibfnamefont
  {L.}~\bibnamefont {Hoepfner}}, \bibinfo {author} {\bibfnamefont
  {S.}~\bibnamefont {Kelterborn}}, \bibinfo {author} {\bibfnamefont
  {M.}~\bibnamefont {Scholz}}, \bibinfo {author} {\bibfnamefont
  {J.}~\bibnamefont {Beißel}}, \bibinfo {author} {\bibfnamefont
  {P.}~\bibnamefont {Hegemann}}, \bibinfo {author} {\bibfnamefont
  {O.}~\bibnamefont {Bäumchen}}, \bibinfo {author} {\bibfnamefont {L.-N.}\
  \bibnamefont {Liu}}, \bibinfo {author} {\bibfnamefont {K.}~\bibnamefont
  {Huang}},\ and\ \bibinfo {author} {\bibfnamefont {M.}~\bibnamefont
  {Hippler}},\ }\bibfield  {title} {\bibinfo {title} {Altered \textit{N}-glycan
  composition impacts flagella-mediated adhesion in \textit{Chlamydomonas
  reinhardtii}},\ }\href {https://doi.org/10.7554/eLife.58805} {\bibfield
  {journal} {\bibinfo  {journal} {eLife}\ }\textbf {\bibinfo {volume} {9}},\
  \bibinfo {pages} {e58805} (\bibinfo {year} {2020})}\BibitemShut {NoStop}%
\bibitem [{Note1()}]{Note1}%
  \BibitemOpen
  \bibinfo {note} {Cell tracking code is available at \protect \url
  {https://doi.org/10.5281/zenodo.4449791}.}\BibitemShut {Stop}%
\bibitem [{\citenamefont {Collingridge}\ \emph {et~al.}(2013)\citenamefont
  {Collingridge}, \citenamefont {Brownlee},\ and\ \citenamefont
  {Wheeler}}]{collingridge2013compartmentalized}%
  \BibitemOpen
  \bibfield  {author} {\bibinfo {author} {\bibfnamefont {P.}~\bibnamefont
  {Collingridge}}, \bibinfo {author} {\bibfnamefont {C.}~\bibnamefont
  {Brownlee}},\ and\ \bibinfo {author} {\bibfnamefont {G.~L.}\ \bibnamefont
  {Wheeler}},\ }\bibfield  {title} {\bibinfo {title} {Compartmentalized calcium
  signaling in cilia regulates intraflagellar transport},\ }\href@noop {}
  {\bibfield  {journal} {\bibinfo  {journal} {Curr. Biol.}\ }\textbf {\bibinfo
  {volume} {23}},\ \bibinfo {pages} {2311} (\bibinfo {year}
  {2013})}\BibitemShut {NoStop}%
\bibitem [{\citenamefont {Fort}\ \emph {et~al.}(2021)\citenamefont {Fort},
  \citenamefont {Collingridge}, \citenamefont {Brownlee},\ and\ \citenamefont
  {Wheeler}}]{fort2021ca2+}%
  \BibitemOpen
  \bibfield  {author} {\bibinfo {author} {\bibfnamefont {C.}~\bibnamefont
  {Fort}}, \bibinfo {author} {\bibfnamefont {P.}~\bibnamefont {Collingridge}},
  \bibinfo {author} {\bibfnamefont {C.}~\bibnamefont {Brownlee}},\ and\
  \bibinfo {author} {\bibfnamefont {G.}~\bibnamefont {Wheeler}},\ }\bibfield
  {title} {\bibinfo {title} {Ca2+ elevations disrupt interactions between
  intraflagellar transport and the flagella membrane in chlamydomonas},\
  }\href@noop {} {\bibfield  {journal} {\bibinfo  {journal} {J. Cell Sci.}\
  }\textbf {\bibinfo {volume} {134}},\ \bibinfo {pages} {jcs253492} (\bibinfo
  {year} {2021})}\BibitemShut {NoStop}%
\bibitem [{\citenamefont {Kural}\ \emph {et~al.}(2005)\citenamefont {Kural},
  \citenamefont {Kim}, \citenamefont {Syed}, \citenamefont {Goshima},
  \citenamefont {Gelfand},\ and\ \citenamefont {Selvin}}]{kural2005kinesin}%
  \BibitemOpen
  \bibfield  {author} {\bibinfo {author} {\bibfnamefont {C.}~\bibnamefont
  {Kural}}, \bibinfo {author} {\bibfnamefont {H.}~\bibnamefont {Kim}}, \bibinfo
  {author} {\bibfnamefont {S.}~\bibnamefont {Syed}}, \bibinfo {author}
  {\bibfnamefont {G.}~\bibnamefont {Goshima}}, \bibinfo {author} {\bibfnamefont
  {V.~I.}\ \bibnamefont {Gelfand}},\ and\ \bibinfo {author} {\bibfnamefont
  {P.~R.}\ \bibnamefont {Selvin}},\ }\bibfield  {title} {\bibinfo {title}
  {Kinesin and dynein move a peroxisome in vivo: a tug-of-war or coordinated
  movement?},\ }\href@noop {} {\bibfield  {journal} {\bibinfo  {journal}
  {Science}\ }\textbf {\bibinfo {volume} {308}},\ \bibinfo {pages} {1469}
  (\bibinfo {year} {2005})}\BibitemShut {NoStop}%
\bibitem [{Note2()}]{Note2}%
  \BibitemOpen
  \bibinfo {note} {See Supplemental Material at [URL] for details of the
  theoretical models, simulation parameters and additional experimental
  data.}\BibitemShut {Stop}%
\bibitem [{\citenamefont {Mecke}(2000)}]{Mecke2000}%
  \BibitemOpen
  \bibfield  {author} {\bibinfo {author} {\bibfnamefont {K.~R.}\ \bibnamefont
  {Mecke}},\ }\bibfield  {title} {\bibinfo {title} {Additivity, convexity, and
  beyond: Applications of minkowski functionals in statistical physics},\ }in\
  \href@noop {} {\emph {\bibinfo {booktitle} {Statistical Physics and Spatial
  Statistics}}},\ \bibinfo {editor} {edited by\ \bibinfo {editor}
  {\bibfnamefont {K.~R.}\ \bibnamefont {Mecke}}\ and\ \bibinfo {editor}
  {\bibfnamefont {D.}~\bibnamefont {Stoyan}}}\ (\bibinfo  {publisher} {Springer
  Berlin Heidelberg},\ \bibinfo {address} {Berlin, Heidelberg},\ \bibinfo
  {year} {2000})\ pp.\ \bibinfo {pages} {111--184}\BibitemShut {NoStop}%
\bibitem [{\citenamefont {Herminghaus}\ \emph {et~al.}(1998)\citenamefont
  {Herminghaus}, \citenamefont {Jacobs}, \citenamefont {Mecke}, \citenamefont
  {Bischof}, \citenamefont {Fery}, \citenamefont {Ibn-Elhaj},\ and\
  \citenamefont {Schlagowski}}]{Herminghaus1998}%
  \BibitemOpen
  \bibfield  {author} {\bibinfo {author} {\bibfnamefont {S.}~\bibnamefont
  {Herminghaus}}, \bibinfo {author} {\bibfnamefont {K.}~\bibnamefont {Jacobs}},
  \bibinfo {author} {\bibfnamefont {K.}~\bibnamefont {Mecke}}, \bibinfo
  {author} {\bibfnamefont {J.}~\bibnamefont {Bischof}}, \bibinfo {author}
  {\bibfnamefont {A.}~\bibnamefont {Fery}}, \bibinfo {author} {\bibfnamefont
  {M.}~\bibnamefont {Ibn-Elhaj}},\ and\ \bibinfo {author} {\bibfnamefont
  {S.}~\bibnamefont {Schlagowski}},\ }\bibfield  {title} {\bibinfo {title}
  {Spinodal dewetting in liquid crystal and liquid metal films},\ }\href
  {https://doi.org/10.1126/science.282.5390.916} {\bibfield  {journal}
  {\bibinfo  {journal} {Science}\ }\textbf {\bibinfo {volume} {282}},\ \bibinfo
  {pages} {916} (\bibinfo {year} {1998})}\BibitemShut {NoStop}%
\bibitem [{\citenamefont {Hoffmann}\ \emph {et~al.}(2006)\citenamefont
  {Hoffmann}, \citenamefont {Ebert}, \citenamefont {Likos}, \citenamefont
  {L\"owen},\ and\ \citenamefont {Maret}}]{Hoffmann2006}%
  \BibitemOpen
  \bibfield  {author} {\bibinfo {author} {\bibfnamefont {N.}~\bibnamefont
  {Hoffmann}}, \bibinfo {author} {\bibfnamefont {F.}~\bibnamefont {Ebert}},
  \bibinfo {author} {\bibfnamefont {C.~N.}\ \bibnamefont {Likos}}, \bibinfo
  {author} {\bibfnamefont {H.}~\bibnamefont {L\"owen}},\ and\ \bibinfo {author}
  {\bibfnamefont {G.}~\bibnamefont {Maret}},\ }\bibfield  {title} {\bibinfo
  {title} {Partial clustering in binary two-dimensional colloidal
  suspensions},\ }\href {https://doi.org/10.1103/PhysRevLett.97.078301}
  {\bibfield  {journal} {\bibinfo  {journal} {Phys. Rev. Lett.}\ }\textbf
  {\bibinfo {volume} {97}},\ \bibinfo {pages} {078301} (\bibinfo {year}
  {2006})}\BibitemShut {NoStop}%
\bibitem [{\citenamefont {Vogel}\ \emph {et~al.}(2010)\citenamefont {Vogel},
  \citenamefont {Weller},\ and\ \citenamefont {Schlüter}}]{vogel2010}%
  \BibitemOpen
  \bibfield  {author} {\bibinfo {author} {\bibfnamefont {H.-J.}\ \bibnamefont
  {Vogel}}, \bibinfo {author} {\bibfnamefont {U.}~\bibnamefont {Weller}},\ and\
  \bibinfo {author} {\bibfnamefont {S.}~\bibnamefont {Schlüter}},\ }\bibfield
  {title} {\bibinfo {title} {Quantification of soil structure based on
  minkowski functions},\ }\href@noop {} {\bibfield  {journal} {\bibinfo
  {journal} {Comput. Geosci.}\ }\textbf {\bibinfo {volume} {36}},\ \bibinfo
  {pages} {1236} (\bibinfo {year} {2010})}\BibitemShut {NoStop}%
\bibitem [{\citenamefont {Mecke}\ \emph {et~al.}(1994)\citenamefont {Mecke},
  \citenamefont {Buchert},\ and\ \citenamefont {Wagner}}]{Mecke1994}%
  \BibitemOpen
  \bibfield  {author} {\bibinfo {author} {\bibfnamefont {K.~R.}\ \bibnamefont
  {Mecke}}, \bibinfo {author} {\bibfnamefont {T.}~\bibnamefont {Buchert}},\
  and\ \bibinfo {author} {\bibfnamefont {H.}~\bibnamefont {Wagner}},\
  }\bibfield  {title} {\bibinfo {title} {{Robust morphological measures for
  large scale structure in the universe}},\ }\href@noop {} {\bibfield
  {journal} {\bibinfo  {journal} {Astron. Astrophys.}\ }\textbf {\bibinfo
  {volume} {288}},\ \bibinfo {pages} {697} (\bibinfo {year}
  {1994})}\BibitemShut {NoStop}%
\bibitem [{\citenamefont {Mantz}\ \emph {et~al.}(2008)\citenamefont {Mantz},
  \citenamefont {Jacobs},\ and\ \citenamefont {Mecke}}]{Mantz2008}%
  \BibitemOpen
  \bibfield  {author} {\bibinfo {author} {\bibfnamefont {H.}~\bibnamefont
  {Mantz}}, \bibinfo {author} {\bibfnamefont {K.}~\bibnamefont {Jacobs}},\ and\
  \bibinfo {author} {\bibfnamefont {K.}~\bibnamefont {Mecke}},\ }\bibfield
  {title} {\bibinfo {title} {Utilizing {M}inkowski functionals for image
  analysis: a marching square algorithm},\ }\href
  {https://doi.org/10.1088/1742-5468/2008/12/p12015} {\bibfield  {journal}
  {\bibinfo  {journal} {J. Stat. Mech.: Theory Exp.}\ }\textbf {\bibinfo
  {volume} {2008}}\bibinfo  {number} { (12)},\ \bibinfo {pages}
  {P12015}}\BibitemShut {NoStop}%
\bibitem [{\citenamefont {Allen}\ and\ \citenamefont
  {Tildesley}(2017)}]{allen2017}%
  \BibitemOpen
\bibfield  {number} {  }\bibfield  {author} {\bibinfo {author} {\bibfnamefont
  {M.~P.}\ \bibnamefont {Allen}}\ and\ \bibinfo {author} {\bibfnamefont
  {D.~J.}\ \bibnamefont {Tildesley}},\ }\href@noop {} {\emph {\bibinfo {title}
  {Computer simulation of liquids}}}\ (\bibinfo  {publisher} {Oxford university
  press},\ \bibinfo {year} {2017})\BibitemShut {NoStop}%
\bibitem [{\citenamefont {Fujiu}\ \emph {et~al.}(2011)\citenamefont {Fujiu},
  \citenamefont {Nakayama}, \citenamefont {Iida}, \citenamefont {Sokabe},\ and\
  \citenamefont {Yoshimura}}]{fujiu2011mechanoreception}%
  \BibitemOpen
  \bibfield  {author} {\bibinfo {author} {\bibfnamefont {K.}~\bibnamefont
  {Fujiu}}, \bibinfo {author} {\bibfnamefont {Y.}~\bibnamefont {Nakayama}},
  \bibinfo {author} {\bibfnamefont {H.}~\bibnamefont {Iida}}, \bibinfo {author}
  {\bibfnamefont {M.}~\bibnamefont {Sokabe}},\ and\ \bibinfo {author}
  {\bibfnamefont {K.}~\bibnamefont {Yoshimura}},\ }\bibfield  {title} {\bibinfo
  {title} {Mechanoreception in motile flagella of {C}hlamydomonas},\
  }\href@noop {} {\bibfield  {journal} {\bibinfo  {journal} {Nat. Cell Biol.}\
  }\textbf {\bibinfo {volume} {13}},\ \bibinfo {pages} {630} (\bibinfo {year}
  {2011})}\BibitemShut {NoStop}%
\bibitem [{\citenamefont {Hornischer}\ \emph {et~al.}(2019)\citenamefont
  {Hornischer}, \citenamefont {Herminghaus},\ and\ \citenamefont
  {Mazza}}]{hornischer2019structural}%
  \BibitemOpen
  \bibfield  {author} {\bibinfo {author} {\bibfnamefont {H.}~\bibnamefont
  {Hornischer}}, \bibinfo {author} {\bibfnamefont {S.}~\bibnamefont
  {Herminghaus}},\ and\ \bibinfo {author} {\bibfnamefont {M.~G.}\ \bibnamefont
  {Mazza}},\ }\bibfield  {title} {\bibinfo {title} {Structural transition in
  the collective behavior of cognitive agents},\ }\href@noop {} {\bibfield
  {journal} {\bibinfo  {journal} {Sci. Rep.}\ }\textbf {\bibinfo {volume}
  {9}},\ \bibinfo {pages} {1} (\bibinfo {year} {2019})}\BibitemShut {NoStop}%
\bibitem [{\citenamefont {Hornischer}\ \emph {et~al.}(2022)\citenamefont
  {Hornischer}, \citenamefont {Pritz}, \citenamefont {Pritz}, \citenamefont
  {Mazza},\ and\ \citenamefont {Boos}}]{hornischer2022}%
  \BibitemOpen
  \bibfield  {author} {\bibinfo {author} {\bibfnamefont {H.}~\bibnamefont
  {Hornischer}}, \bibinfo {author} {\bibfnamefont {P.~J.}\ \bibnamefont
  {Pritz}}, \bibinfo {author} {\bibfnamefont {J.}~\bibnamefont {Pritz}},
  \bibinfo {author} {\bibfnamefont {M.~G.}\ \bibnamefont {Mazza}},\ and\
  \bibinfo {author} {\bibfnamefont {M.}~\bibnamefont {Boos}},\ }\bibfield
  {title} {\bibinfo {title} {Modeling of human group coordination},\
  }\href@noop {} {\bibfield  {journal} {\bibinfo  {journal} {Phys. Rev. Res.}\
  }\textbf {\bibinfo {volume} {4}},\ \bibinfo {pages} {023037} (\bibinfo {year}
  {2022})}\BibitemShut {NoStop}%
\bibitem [{\citenamefont {Wissner-Gross}\ and\ \citenamefont
  {Freer}(2013)}]{wissnerPRL2013}%
  \BibitemOpen
  \bibfield  {author} {\bibinfo {author} {\bibfnamefont {A.~D.}\ \bibnamefont
  {Wissner-Gross}}\ and\ \bibinfo {author} {\bibfnamefont {C.~E.}\ \bibnamefont
  {Freer}},\ }\bibfield  {title} {\bibinfo {title} {Causal entropic forces},\
  }\href {https://doi.org/10.1103/PhysRevLett.110.168702} {\bibfield  {journal}
  {\bibinfo  {journal} {Phys. Rev. Lett.}\ }\textbf {\bibinfo {volume} {110}},\
  \bibinfo {pages} {168702} (\bibinfo {year} {2013})}\BibitemShut {NoStop}%
\bibitem [{\citenamefont {van~der Marck}(1997)}]{van1997percolation}%
  \BibitemOpen
  \bibfield  {author} {\bibinfo {author} {\bibfnamefont {S.~C.}\ \bibnamefont
  {van~der Marck}},\ }\bibfield  {title} {\bibinfo {title} {Percolation
  thresholds and universal formulas},\ }\href@noop {} {\bibfield  {journal}
  {\bibinfo  {journal} {Phys. Rev. E}\ }\textbf {\bibinfo {volume} {55}},\
  \bibinfo {pages} {1514} (\bibinfo {year} {1997})}\BibitemShut {NoStop}%
\bibitem [{\citenamefont {Meyer}\ \emph {et~al.}(2010)\citenamefont {Meyer},
  \citenamefont {Song}, \citenamefont {Jin}, \citenamefont {Wang},\ and\
  \citenamefont {Makse}}]{meyer2010jamming}%
  \BibitemOpen
  \bibfield  {author} {\bibinfo {author} {\bibfnamefont {S.}~\bibnamefont
  {Meyer}}, \bibinfo {author} {\bibfnamefont {C.}~\bibnamefont {Song}},
  \bibinfo {author} {\bibfnamefont {Y.}~\bibnamefont {Jin}}, \bibinfo {author}
  {\bibfnamefont {K.}~\bibnamefont {Wang}},\ and\ \bibinfo {author}
  {\bibfnamefont {H.~A.}\ \bibnamefont {Makse}},\ }\bibfield  {title} {\bibinfo
  {title} {Jamming in two-dimensional packings},\ }\href@noop {} {\bibfield
  {journal} {\bibinfo  {journal} {Physica A}\ }\textbf {\bibinfo {volume}
  {389}},\ \bibinfo {pages} {5137} (\bibinfo {year} {2010})}\BibitemShut
  {NoStop}%
\end{thebibliography}
\end{document}